\documentstyle[12pt]{article}

\newcounter{popnr}
\def\theequation{\thesection.\arabic{equation}}
\renewcommand{\theequation}{\arabic{section}.\arabic{equation}}
\newcommand{\alpheqn}{\setcounter{popnr}{\value{equation}}
		      \addtocounter{popnr}{1}
		      \setcounter{equation}{0}
\renewcommand{\theequation}{\arabic{section}.\arabic{popnr}\alph{equation}}}
\newcommand{\reseteqn}{\setcounter{equation}{\value{popnr}}
     \renewcommand{\theequation}
     {\arabic{section}.\arabic{equation}}}

\normalbaselines
\catcode `\@=11
\@addtoreset{equation}{section}
\def\theequation{\arabic{section}.\arabic{equation}}
\def\section{\@startsection {section}{1}{\z@}{-3.5ex plus -1ex minus
     -.2ex}{2.3ex plus .2ex}{\normalsize\bf}}
\def\subsection{\@startsection{subsection}{2}{\z@}{-3.25ex plus -1ex minus
 -.2ex}{1.5ex plus .2ex}{\normalsize\bf}}
\def\thebibliography#1{\section*{References\markboth
  {REFERENCES}{REFERENCES}}\list
  {\arabic{enumi}.}{\settowidth\labelwidth{[#1]}\leftmargin\labelwidth
  \advance\leftmargin\labelsep
  \usecounter{enumi}}
  \def\newblock{\hskip .11em plus .33em minus -.07em}
  \sloppy
  \sfcode`\.=1000\relax}
 
\catcode `\@=12

\begin{document}

\noindent

\begin{flushright}
NEIP-99-004\\
gr-qc/9903066 \\
$~$ \\
Int.~J.~Mod.~Phys.~A (in press)
\end{flushright}

\bigskip

\begin{center}
{\bf DISTANCE MEASUREMENT AND $\kappa$-DEFORMED}\\
{\bf PROPAGATION OF LIGHT AND HEAVY PROBES}
\end{center}
\hspace*{1in}
\begin{center}
Giovanni Amelino-Camelia$^{a, }$\footnote{Supported by a grant of
the Swiss National Science
Foundation}, Jerzy Lukierski$^{b}$
and Anatol Nowicki$^{c, }$\footnote{Partially supported by KBN grant
2P03B130.12}
\end{center}
\begin{center}
\begin{minipage}{13cm}
 $^{a}$ Institut de Physique, Universit\'e de Neuch\^atel,\\
\makebox[3mm]{ } rue Breguet 1, CH-2000 Neuch\^atel, Switzerland \\
 $^{b}$ Institute for Theoretical Physics, University of Wroc{\l}aw,\\
\makebox[3mm]{ } pl. M. Borna 9, 50-204 Wroc{\l}aw, Poland \\
 $^{c}$ Institute of Physics, Pedagogical University,\\
\makebox[3mm]{ } pl. S\l{}owia\'nski 6, 65-029 Zielona G\'ora, Poland\\
\makebox[3mm]{ }
\end{minipage}
\end{center}


\begin{abstract}
\noindent
We investigate the implications for the measurability of
distances of a covariant dimensionful ``$\kappa$''
deformation of $D=4$ relativistic symmetries,
with quantum time coordinate and modified Heisenberg algebra.
We show that the structure of the deformed 
mass-shell condition has significant implications for
measurement procedures relying on light probes,
whereas in the case of heavy probes
the most sizeable effect is due to
the nontrivial commutation relation
between three-momenta and quantum time coordinate.
We argue that these findings might indicate
that $\kappa$-Poincar\'e symmetries capture
some aspects of the physics of
the Quantum-Gravity vacuum.

\end{abstract}

\baselineskip=13pt

\section{\hspace{-4mm}.\hspace{2mm}Introduction}

\hspace*{1cm}
One of the aspects of Quantum Gravity that has been most
actively investigated is the possibility that there be
a minimum uncertainty for the measurement of distances.
The simplest and best understood proposal is the one of
a ``minimum length''
\begin{equation}\label{1.2}
\Delta	x \ge L_{min} \, ,
\end{equation}
which fits well the expectation of certain
studies \cite{padma,dopp} of measurability
in quantum gravity. Eq.~(\ref{1.2}) might also hold
in critical sting theory as suggested
by analyses of string collisions at Planckian
energies, which were found \cite{vene} to be characterized
by the following modified uncertainty relation
\begin{equation}\label{1.1}
\Delta	x \geq \frac{\hbar}{\Delta  p} + \alpha G \Delta  p\, ,
\end{equation}
where $G=c^2 l_p^2/\hbar$ is the gravitational coupling (Newton) constant,
$c$ and $l_p$ are the speed-of-light and Planck-length
constants respectively,
and $\alpha$ is a constant related to the
string tension (Regge slope).
Clearly Eq.~(\ref{1.1}) implies $L_{min} \sim \sqrt{\hbar \alpha G}$.

While modified uncertainty relations are not necessarily
associated to quantum groups~\cite{dopp}, it is interesting
that quantum-group descriptions are often available~\cite{majbook,magg,kempf};
in particular,
in Ref.~\cite{kempf}
it was shown that the relations (\ref{1.2})-(\ref{1.1})
can be associated to $SU_q(n)$ covariance.

The minimum-length scenario (\ref{1.2}) 
would already require that the
conceptual framework of Quantum Gravity 
be significantly different from the
one of ``ordinary'' (non-gravitational) 
Quantum Mechanics\footnote{In particular,
in Ref.~\cite{ahlu} it was shown that 
when gravitational effects are taken into account in 
a (quantum) measurement process
then the masses of the probes used in the measurement
induce a change in the
space-time metric and this 
is associated to the emergence of nonlocality.
The nature of this 
gravitationally-induced nonlocality 
suggests \cite{ahlu} a modification of the 
fundamental commutators.}.
However, as emphasized in Ref.~\cite{gacmpla},
even more dramatic consequences for the
measurability of distances appear to emerge from the analysis
of additional contributions to the uncertainty relations
which are associated to the fact that the
limit of ``classical'' (infinitely massive) devices
might not be accessible in Quantum Gravity.
In particular, the analysis in Ref.~\cite{gacmpla}
has led to the proposal of an alternative to the bound (\ref{1.2}).
By taking into account both the quantum nature of the agents
involved in the measurement and the gravitational
effects associated to the devices,
one finds \cite{gacmpla} that the measurability of distances
is bound by a quantity that (as needed for the decoherence
mechanism discussed in Ref. \cite{karo})
grows with the time required
by the measurement procedure
\begin{equation}\label{qgboundgac}
\min \left[ \Delta L \right] \sim l_p \sqrt{ c T \over s} \sim
l_p\sqrt{ L \over s }
~,
\end{equation}
where $L$ is the distance being measured,
$s$ is a length scale characterizing
the spatial extension of the devices ({\it e.g.}, clocks)
used in the measurement, $T$ is the
time needed to complete the procedure of measuring $L$,
and on the right-hand-side we used the fact that, assuming the
measurement procedure uses massless probes,
one has typically $T \sim L$.
Notice that for all acceptable values \cite{gacmpla}
of $s$ ({\it i.e.} $s < L$) the bound
(\ref{qgboundgac}) is more stringent than (\ref{1.2});
this is a direct consequence of the fact that the analyses
leading to (\ref{1.2}) had implicitly relied on the availability of
ideal classical devices in the measurement procedure.

While, as mentioned above, critical string theories provide a framework
for the bound (\ref{1.2}), it appears that noncritical string theories
might provide a framework for the bound (\ref{qgboundgac}). In particular,
within ``Liouville'' noncritical string theories, in which the
target time is identified with the Liouville mode \cite{emn},
the nature of the dynamics of the light probes
exchanged in a typical procedure of measurement of a distance
was shown \cite{aemn1} to lead to a
measurability bound of type (\ref{qgboundgac}).

An interesting problem is the one of finding a quantum-group
(and quantum-Lie-algebra)  framework
for (\ref{qgboundgac}),
just like Ref.\cite{kempf} has provided
a quantum-group framework for (\ref{1.2})-(\ref{1.1}).
The notion of quantum group as a Hopf algebra permits to consider
deformed symmetries; in fact, the Hopf algebra	axioms
provide simultaneously an algebraic generalization of the definition
of Lie group as well as of Lie algebra.
As exemplified by the formulae
in the following section,
the phase space containing the
coordinate and momentum sectors can be described in
the quantum-deformed case
as a semidirect product of two
dual Hopf algebras describing the coordinates
and momentum sectors.
Such a definition of quantum phase space has been first
proposed by Majid \cite{majbook},
and it is endoved with the property that in the undeformed case
(coordinates and momenta described by Abelian Hopf algebra
with primitive
coproducts) one obtains the standard quantum mechanical
Heisenberg commutation relations.\footnote{In the literature
sometimes
the semidirect product construction for two dual Hopf algebras
describing respectively quantum Lie group and quantum Lie algebra
is called
``Heisenberg double'' (see, {\it e.g.}, Ref. \cite{fadde}).}
The so-called $\kappa$-deformations \cite{lnr1,lnr2,mr,lrz}
provide an example of this type
of quantum deformations of relativistic symmetries, and
one of us recently argued \cite{kpoinpap} that
$\kappa$-deformed symmetries might provide
an algebraic abstraction of the
measurability bound (\ref{qgboundgac}).
The analysis reported in \cite{kpoinpap}
was somewhat preliminary since only
the coordinate sector was considered,
but the bound (\ref{qgboundgac}) emerged rather compellingly,
as a direct consequence of the noncommuting space-time coordinates
of $\kappa$-deformed Minkowski space.
Encouraged by the findings of Ref.\cite{kpoinpap},
in this paper we explore further the relation
between $\kappa$-Poincar\'e
and (\ref{qgboundgac}); specifically, we extend
the analysis of  Ref.\cite{kpoinpap}
from the confines of the space-time coordinate sector
to the full structure of the $\kappa$-deformed phase space.
We primarily consider the $\kappa$-deformed Poincar\'e symmetries 
in the bicrossproduct basis \cite{majbook,mr},
which appears to be a very natural framework for the
quantum deformations of semidirect product algebras,
and outside of the coordinate sector we
identify two structures which could affect the analysis
of Ref.\cite{kpoinpap}:
the $\kappa$-deformed mass-shell condition,
which is associated to the Casimir and
suggests a modification of the propagation of the
light probes exchanged during measurement,
and the nontrivial commutation relation
between three-momenta and quantum time coordinate, which
we find to have important consequences
for the analysis of the propagation of
heavy probes exchanged during measurement.
As discussed below,
our analysis uncovers new nonnegligible
contributions to the bound
on the measurability of distances.
These contributions
are however comparable to the one identified
in Ref.\cite{kpoinpap}, and therefore
the order of magnitude of the effect
discussed in Ref.\cite{kpoinpap} is confirmed by our analysis.
These findings provide additional evidence of a relation
between $\kappa$-Poincar\'e and the bound (\ref{qgboundgac}),
and thereby contribute to the development of a physical 
interpretation of this class of deformations.
We hope that this will provide further motivation
for experimentalists to 
investigate the theoretical framework here advocated, 
especially exploiting the recent remarkable 
discoveries \cite{grbnew} in the phenomenology of
gamma-ray bursts that
allow \cite{grbgac} a direct test\footnote{Previous
analyses attempting to bound the dimensionful parameter $\kappa$
only probed values of $\kappa$ that were several orderds of magnitude
below the Planck scale (see, {\it e.g.}, Ref.~\cite{aemn1,domo,bowjar}),
but now that we have
definitive evidence that gamma-ray bursts are at cosmological distances
we can expect \cite{grbgac} to probe values of $\kappa$ all the way
up to the Planck scale.}
of some of the predictions of the $\kappa$ deformations
of Poincar\'e symmetries.

\section{$\kappa$-deformed quantum relativistic phase space}

The standard form of the covariant fourdimensional Heisenberg commutation
relations, describing quantum-mechanical covariant phase space looks
as follows:
\begin{equation}\label{2.1}
[x_\mu,p_\nu] = i\hbar g_{\mu\nu}\,,	 \qquad g_{\mu\nu} =
\mathop{\rm diag} (-1,1,1,1).
\end{equation}
The space-time coordinates $x_\mu$
($\mu=0, 1, 2, 3$)
can be identified with the translation sector of the Poincar\'e
group, and the fourmomenta $p_\mu$ ($\mu=0, 1, 2, 3$) are given by
the translation generators of the Poincar\'e algebra.
In considering
quantum deformations of relativistic symmetries as describing
the modification of space-time
structure one is lead to the study of the
possible quantum Poincar\'e
groups.\footnote{We take into consideration here only the
genuine 10-generator quantum deformations of $D=4$ Poincar\'e symmetries.
In particular, the ``standard'' $q$-deformations are not considered.
These $q$-deformations require adding an
eleventh (dilatation) generator,
{\it i.e.} one deals with the dilatation extended Poincar\'e
algebra \cite{majpap}.
In such a case the corresponding quantum phase space
is much more complicated (see, {\it e.g.}, \cite{wessnew}), and the
deformation parameter is dimensionless, rendering difficult
the physical separation between the ordinary regime
of commutative space-time coordinates and the short-distance regime
in which non-commutativity sets in.}
The classification of quantum
deformations  of $D=4$ Poincar\'e groups in the
framework of Hopf algebras was given by
Podle\'s and Woronowicz (\cite{pod1}; see also \cite{pod2})
and provides the most general
class of noncommutative space-time coordinates $\hat x_\nu$
allowed by the quantum-group formalism.
If we assume that the quantum deformation does
not affect the nonrelativistic kinematics, i.e. we preserve the
nonrelativistic $O(3)$ rotations classical
and $O(3)$ covariance, the
only consistent class of
noncommuting space-time coordinates is described by the relations
of the $\kappa$-deformed Minkowski space
with commuting classical space coordinates.
In order to describe the relativistic phase space
we start with the deformed Hopf algebra
of fourmomenta ${\hat p}_{\mu}$
written as follows
\begin{eqnarray}
[\hat  p_{0}, \hat  p_{k}] &=& 0 \label{2.2} \\
\Delta (\hat  p_0) &=& \hat  p_0 \otimes 1 + 1\otimes \hat  p_0 \nonumber \\
\Delta (\hat  p_k) &=& \hat  p_k \otimes e^{\alpha{\hat p_{0}}} +
e^{\beta{\hat  p_{0}}}\otimes \hat
p_k\label{2.3} \end{eqnarray}
with antipode and counit given by
\begin{equation}\label{2.4}
S(\hat	p_k) = - e^{-(\alpha+\beta)\hat  p_0} \, \hat  p_k
\qquad\qquad
S(\hat	p_0) = -\hat  p_0 \qquad\qquad \epsilon(\hat  p_\mu)=0\,.
\end{equation}

Using the duality relations involving the
fundamental constant $\hbar$
(Planck's constant)
\begin{equation}\label{2.5}
\left<\hat x_\mu, \hat	p_\nu\right> = -i\hbar g_{\mu\nu}\qquad g_{\mu\nu} = (-1,1,1,1)
\end{equation}
we obtain the noncommutative deformed configuration space
${\cal{X}}$ as a Hopf algebra with the following algebra and
coalgebra structure \bigskip
\begin{eqnarray}\label{2.6}
[\hat x_{0}, \hat x_{k}] &=& {i\hbar}(\beta-\alpha)\hat x_k\,,\qquad\qquad
[\hat x_{k}, \hat x_{l}] = 0 \,,
\end{eqnarray}

\alpheqn
\begin{eqnarray}
\Delta (\hat x_\mu ) &=& \hat x_{\mu}\otimes 1 + 1\otimes \hat x_{\mu}\,,
\\ S(\hat x_{\mu}) &=& -\hat x_{\mu} \,, \qquad\qquad
\epsilon(\hat x_{\mu}) = 0 \,.
\end{eqnarray}
\reseteqn
The deformed phase space can be considered as the
vector space ${\cal{X}}\otimes
{\cal{P}}$ with the product (see \cite{majbook})
\begin{equation}\label{2.8}
(x\otimes p)(\tilde{x}\otimes \tilde{p})=x(p_{(1)}\triangleright
\tilde{x})\otimes p_{(2)}\tilde{p}
\end{equation}
where left action is given by
\begin{equation}\label{2.9}
p\triangleright x=\left<p,x_{(2)}\right>x_{(1)}
\end{equation}
The product (\ref{2.8}) can be rewritten as the commutators between
coordinates and momenta by using the obvious isomorphism
$x\sim x\otimes 1$, $p\sim 1\otimes p$.
This procedure	provides the following commutation relations (see also
\cite{ln,now})
\begin{equation}\label{2.10}
\begin{array}{rclrcl}
[\hat x_k,\hat	p_l] &=&i \hbar \delta _{kl}\,e^{\alpha{\hat p_0}}\,,
\qquad&[\hat x_k,\hat  p_0]&=&0\,,\\[3mm]
[\hat x_0,\hat	p_k] &=& -{{i\hbar}\beta} \hat	p_k\,,
\qquad&[\hat x_0,\hat  p_0]&=&-i\hbar\,. \end{array}
\end{equation}
The set of relations $(2.2)$, $(2.6)$ and $(2.10)$
describes the
deformed relativistic quantum phase space.

Introducing the dispersion of the observable
$a$ in quantum mechanical sense by
\begin{equation}\label{2.11}
\Delta (a) \ = \ \sqrt{\left<a^2\right> - \left<a\right>^2}
\end{equation}
we have
\begin{equation}\label{2.12}
\Delta (a)\Delta (b)\geq {1\over 2}|\left<c\right>|\qquad\mbox{ where }\qquad
c=[a,b] \end{equation}
We obtain deformed uncertainty relations in the form
\alpheqn
\begin{eqnarray}\label{2.13}
\Delta	\hat x_0 \Delta  \hat x_k &\geq& {\hbar\over 2}
|(\beta-\alpha)||\left<\hat x_k\right>|\\
\Delta	\hat  p_k  \Delta  \hat x_l &\geq& {\hbar\over 2}\delta
_{kl}\left<e^{\alpha {\hat p_0}}\right>\\
\Delta	\hat p_0  \Delta  \hat x_0 &\geq& {\hbar\over 2}\\
\Delta	\hat  p_k \Delta  \hat x_0 &\geq&
{\hbar\over 2}|\left<\beta \hat  p_k\right>|
\end{eqnarray}
\reseteqn
Depending on a choice of the parameters $\alpha$ and $\beta$ we can distinguish
the following cases ($c$-speed of light, $\kappa$-(mass like) deformation
parameter):\bigskip\\
i) $\alpha=\beta=0$\quad;\quad standard form of nondeformed covariant phase
space,\\
ii) $\alpha=\beta$\quad;\quad trivially deformed phase space with commuting
configuration space,\\
iii) $\alpha=-\beta={1\over 2\kappa c}$\quad;\quad $\kappa$-deformed phase
space in the standard basis (see \cite{an}),\\
iv) $\alpha=0, \beta=-{1\over \kappa c}$\quad;\quad $\kappa$-deformed phase
space in the bicrossproduct basis (see \cite{mr}),\\
v) $\alpha={1\over \kappa c}, \beta=0$\quad;\quad $\kappa$-deformed phase space
in the bicrossproduct basis (see \cite{lrz}),\\
vi) $\alpha=0, \beta={1\over \kappa c}$\quad;\quad $\kappa$-deformed phase
space in the bicrossproduct basis (the case (v)) with transposed coproduct.

The Quantum-Gravity arguments advocated in the next sections
make contact with the  bicrossproduct basis, and therefore
(also for definiteness) in the following we focus on the case (vi).
The set of relations $(2.2)$, $(2.6)$ and $(2.10)$ for our choice of
parameters (vi) are the following
\alpheqn
\begin{eqnarray}\label{2.14}
\left[\hat{p}_{0}, \hat{p}_{k}\right] &=&  0  \\
\left[\hat{x}_{0}, \hat{x}_{k}\right] &=& {{i\hbar}\over \kappa c}\hat{x}_k
\,,\qquad \qquad [\hat{x}_{k}, \hat{x}_{l}] = 0 \\
\left[\hat{x}_k,\hat{p}_l\right] &=& i \hbar \delta _{kl}\,,
\qquad\qquad [\hat{x}_k,\hat{p}_0] = 0 \\
\left[\hat{x}_0,\hat{p}_k\right] &=&  -{{i\hbar}\over \kappa c} \hat{p}_k\,,
\qquad\quad [\hat{x}_0,\hat{p}_0] = -i\hbar
\end{eqnarray}
\reseteqn
and are $\kappa$-Poincar\'e
covariant \footnote{The $\kappa$-covariance of the
relations (2.14b) has been shown firstly in Ref.~\cite{mr}.
The $\kappa$-covariance of the whole quantum $\kappa$-deformed Heisenberg
algebra follows from the general properties of the semidirect product,
defined by the relations (2.14b) and (2.14c-d).
(see, {\it e.g.}, \cite{schnew})}.

The modified covariant Heisenberg
uncertainty relations follow
from the relations $(2.14)$,
therefore we obtain $\kappa$-deformed uncertainty relations
\alpheqn
\begin{eqnarray}\label{2.15}
\Delta	\hat t \Delta  \hat x_k &\geq& {\hbar\over 2\kappa
c^2}|\left<\hat x_k\right>|=\frac12\frac{l_\kappa}{c}\, |\left<\hat x_k\right>|
\,,\\
\Delta	\hat  p_k  \Delta  \hat x_l &\geq& {1\over 2}\hbar\delta _{kl}
\,,\\ \Delta  \hat E  \Delta  \hat t &\geq& {1\over 2}\hbar
\,, \\ \Delta  \hat  p_k \Delta  \hat t&\geq& {\hbar\over
2\kappa c^2}|\left<\hat  p_k\right>|=\frac12\frac{l_\kappa}{c}\,
\left|\left<\hat  p_k\right>\right|\,.
\end{eqnarray}
\reseteqn
where $l_\kappa=\frac{\hbar}{\kappa c}$ describes the fundamental
length at which the time variable should already be considered
noncommutative.
In comparison with the discussion
in Ref.\cite{kpoinpap}, which only considered
the coordinate sector,
the significant new element that emerged in our present analysis
is the relation (2.15d).
Interestingly, multiplying the three relations
(2.15a), (2.15b) and (2.15d)
one obtains
\begin{eqnarray}
(\Delta  \hat t )^2 (\Delta  \hat x_l \Delta  \hat  p_l)^2 &\geq&
{\hbar \over 8}
\frac{l_\kappa^2}{c^2} \,
\left| \left<\hat x_l\right> \left<\hat  p_l\right> \right|
\label{2.16}\,
\end{eqnarray}
(where no sum over the index $l$ is to be understood).
This suggests that a wave packet with minimal standard
($\Delta  x \Delta  p$) uncertainty
has the largest uncertainty in the localization of time.
(In ordinary quantum mechanics $l_\kappa = 0$ and there is no
such correlation.)

It is also interesting to consider the relation (2.15d)
under the assumption
that the three-momenta ${\hat p_k}$ can be expressed by a general
formula ${\hat p_i} = {\cal M}(v^2) v_i$, in which case
$\Delta  {\hat p_i} = {\cal M}_{i j} \Delta  v_k$ with
${\cal M}_{i j} = {\cal M} [ \delta _{i j} + 2 v_i v_j (\ln {\cal M})']$.
Then (2.15d) implies
\begin{eqnarray}
\Delta	\hat t \Delta  v_i &\geq&
\frac{l_\kappa}{c} \,
{\cal M}(v) \, {\cal M}_{i j}^{-1} (v) \, v_j
\label{2.17}\,
\end{eqnarray}

Because in part of our measurement analysis we shall consider
light probes,
we now discuss the modification of the kinematics
of $\kappa$-deformed photons.
We shall assume that the generators of the $\kappa$-deformed
Poincar\'e algebra
in bicrossproduct basis describes the ``physical'' generators
of space-time symmetries.
In the bicrossproduct
basis the $\kappa$-deformed mass Casimir takes the form
\begin{equation}\label{2.18}
C_2^\kappa = {1\over c^2}\vec P^2 e^{-\frac{P_0}{\kappa c}}
- (2\kappa \sinh
\frac{P_0}{2\kappa c} )^2 = -M^2\,,
\end{equation}
where $P_\mu$ are the generators of space-time translations and
$M$ denotes the $\kappa$-invariant mass parameter.
For $M=0$ ($\kappa$-deformed photons) from (2.18)
one obtains that
(we identify $P_\mu \equiv {\hat p}_\mu$)
\begin{equation}\label{2.19}
{\hat p}_0=\kappa c \ln (1+\frac{|\vec {\hat p}|}{\kappa c })
= |\vec {\hat p}|
-\frac{|\vec {\hat p}|^2}{2 \kappa c} + O(\frac1{\kappa^2})
\end{equation}
and in particular the velocity
formula for massless $\kappa$-deformed quanta
looks as follows\footnote{The relation (2.20a)
is valid as a consequence of the Hamiltonian equation of motion
${\dot x}_i = \partial H/ \partial p_i
- (x_i/\kappa) \partial H/ \partial x_0$.
[See Ref.~\cite{lrz}, Eq.~(4.22).]
For the $\kappa$-photon here considered,
since $H=H(p_i)$, the velocities are classical ($[v_i,v_j]=0$).}
($E=c {\hat p}_0$)
\alpheqn
\begin{equation}\label{2.20a}
v_i = \frac{\partial E}{\partial {\hat p}_i} = \frac{c}{1+
\frac{|\vec {\hat p}|}{\kappa c}} \frac{{\hat p}_i}{|\vec {\hat p}|}
\end{equation}
or
\begin{equation}\label{2.20b}
v=|\vec{v}|=\frac {c}{1+\frac{|\vec {\hat p}|}{\kappa c}} = c-
\frac{|\vec {\hat p}|}{\kappa} +O(\frac 1 {\kappa^2})
\end{equation}
\reseteqn
The inverse formula, which can be inserted in (2.17),
looks as follows
\begin{equation}\label{2.21}
{\hat p}_i=\kappa {c \over v} ({c \over v} -1) v_i
\end{equation}
and it is linear in the deformation parameter $\kappa$.

\section{Measurement of distance and covariant $\kappa$-deformed
phase space}
In this section we analyze the measurement of the distance $L$
between two bodies as it results from a plausible physical
interpretation of the uncertainty
relations (2.15a)-(2.15d).
Like the related studies \cite{gacmpla,karo,aemn1}
we consider the procedure of measurement of distances
set out by Wigner \cite{wign},
which relies on the exchange of a
probe/signal between the bodies.
The distance is therefore measured as $L = v \, T/2$,
where $v$ is the velocity of the probe and $T$ is the time
(being measured by a clock)
spent by the probe to go from one body to the other and return.
In general the quantum mechanical nature of the agents intervening
in the experiment introduces uncertainties in the measurement of $L$,
and in particular one finds that~\footnote{Of course
there are other contributions to $\Delta  L$ ({\it e.g.}, coming
from the quantum mechanical nature of the other devices used
in the experiment \cite{gacmpla}); however, since they obviously
contribute additively to the total uncertainty
in the measurement of $L$,
these uncertainties could only make stronger
the bound derived in the following.}
\begin{equation}\label{new1}
\Delta L \ge [\Delta  L]_{clock} + [\Delta  L]_{probe}
~,
\end{equation}
{\it i.e.} the uncertainty in the measurement of $L$ receives
of course contributions
that originate from the quantum mechanical nature of the ``clock''
(the timing/triggering device employed in the measurement)
and from the quantum mechanical nature of the
probe exchanged between the bodies.
A significant contribution to $\Delta L_{clock}$
was uncovered in Ref.~\cite{gacmpla}; this results in the relation
\begin{equation}\label{new2}
 [\Delta L]_{clock}  \ge l_p \sqrt{ c T \over s}
~,
\end{equation}
where $s$ is a length scale characterizing
the spatial extension of the clock
({\it e.g.}, the radius of a spherically-symmetric clock)
and $T$ is the
time needed to complete the procedure of measuring $L$
({\it i.e.} $T$ is the time that the clock measures).

Within ordinary quantum mechanics the
quantum mechanical nature of the probe
(while contributing in general to the uncertainty) does not contribute
to the bound on the measurability
of $L$ ({\it i.e.} a suitable measurement
set up can be found so that the quantum mechanical nature of the probe
does not lead to a contribution to $\Delta  L$).
It was shown in Ref.\cite{kpoinpap} that instead the kinematics
of quantum $\kappa$-Minkowski space-time does lead to a
nontrivial $[\Delta  L]_{probe}$, and interestingly
this turns out to be
of the same form of the $[\Delta L]_{clock}$ in (\ref{new2}).
As announced in the Introduction we are interested in
extending the analysis
of Ref.\cite{kpoinpap}
to include structure from the full $\kappa$-deformed phase space.
We are also more general than Ref.\cite{kpoinpap} and
other related work
(see, {\it e.g.}, Ref.\cite{gacmpla,karo,aemn1})
in that we not only consider massless particles
as the probes exchanged
in the Wigner measurement, but we also consider the opposite
limit in which the probes are ultra-heavy.

\subsection{Using a heavy probe}
In general
combining the contribution (\ref{new2}), which originates from
the quantum mechanical nature of the clock,
with uncertainties due to
the quantum mechanical nature of the probe
one finds that
\begin{equation}\label{3.1}
\Delta L \geq l_p \sqrt{ c T \over s} + \Delta	x + v \, \Delta  t + T \Delta  v
\end{equation}
where $\Delta  x$ and $\Delta  t$ are the uncertainties
on the space-time position~\footnote{As implicit in the
terminology here adopted, the Wigner measurement procedure is
essentially one-dimensional, and the only relevant spatial
coordinate is the one along the axis passing through
the bodies whose distance is being measured.}
of the probe at the ``final time'' $T$,
while $\Delta  v$ is the uncertainty on the velocity of the probe.

The first contribution on the right-hand-side of (\ref{3.1})
originates from the quantum mechanical nature of the clock, and
it is interesting to notice that in the case of a heavy probe
the proportionality to $\sqrt{T}$ of that term, which always signals
decoherence effects ({\it e.g.}, the more time goes by, the more the
quantum clock decoheres according to the ideas in Refs.\cite{gacmpla,karo}),
can be turned into a proportionality to $\sqrt{L/v}$, {\it i.e.}
the uncertainty actually diverges in the limit of vanishing velocity
as expected in a context involving decoherence
(small velocities imply large times).

Concerning the contributions on the right-hand-side of (\ref{3.1})
that originate from the quantum mechanical nature of the probe,
it is interesting to observe that in ordinary quantum mechanics
$\Delta  x$, $\Delta  t$ and  $\Delta  v$ are not correlated
and therefore they do not lead to a contribution to the bound on the
measurability of $L$.
However, the $\kappa$ deformation induces correlations
between $\Delta  x$, $\Delta  t$ and  $\Delta  v$.
In particular,
we observe that (2.15a)-(2.15d) imply
(for an ideal heavy/nonrelativistic probe with $p = M v$ and
interpreting the $x$ on the right-hand-side of (2.15a)
as the distance traveled by the probe)
\begin{equation}\label{3.12}
\Delta v
\geq \frac{l_\kappa \, v}{2c \, \Delta t}
\end{equation}
and
\begin{equation}\label{3starbclock}
\Delta	x \geq
\frac{l_\kappa \, L}{2 c \, \Delta  t}
\, .
\end{equation}
This relations together with
the fact that $v \sim L/T$
allow to rewrite (\ref{3.1}) as
\begin{equation}\label{3.1final}
\Delta L \geq l_p \sqrt{ c T \over s} +
\frac{l_\kappa \, L}{2 c \, \Delta  t}
+ {L \over T} \Delta  t
+ \frac{l_\kappa \, L}{2 c \, \Delta  t}
~.
\end{equation}
This uncertainty can be minimized by preparing the probe
in a state with $v \sim c l_p / \sqrt{s l_\kappa}$,
{\it i.e.} $T \sim L  \sqrt{s l_\kappa} /(c l_p)$,
and $\Delta  t \sim \sqrt{ l_\kappa T /  c }$, and this results
in the measurability bound
\begin{equation}\label{3starbis}
min [\Delta L] \sim \sqrt{L l_p \sqrt{{l_\kappa \over s}}}
~.
\end{equation}

The fact that this bound
emerging from our analysis of Wigner measurement
using a heavy probe
manifests the same $\sqrt{L}$ behavior encountered in
the heuristic quantum-gravity
analysis of the clock involved in the measurement
is a rather interesting
aspect of the covariantly $\kappa$-deformed phase space.
In fact,  Eq.(\ref{3.12}),
which reflects the specific structure
of the $\kappa$-deformed commutation relation
between three-momenta and quantum time coordinate,
plays a nontrivial role in establishing that
the $\kappa$-deformed kinematics of the heavy probe
leads to an uncertainty with this $\sqrt{L}$ behavior.

\subsection{Using a massless probe}

Of course, also in the case of a Wigner measurement involving
a massless probe one finds that
\begin{equation}\label{3.1light}
\Delta L \geq l_p \sqrt{ c T \over s} + \Delta	x
+ c \, \Delta  t + T \Delta  v
~,
\end{equation}
and again
the $\kappa$ deformation induces correlations between
$\Delta  x$, $\Delta  t$ and  $\Delta  v$.
In particular,
concerning the correlation between
$\Delta  x$ and
$\Delta  t$ using again (2.15a)
one finds
\begin{equation}\label{3starb}
\Delta	t \geq \frac{\hbar L}{2 \kappa c^2\, \Delta  x}
\, .
\end{equation}
Moreover,  if the probe is massless with
modified velocity\footnote{It
is interesting to notice that $\kappa$-deformed mass-shell condition
and $\kappa$-commutation relation
between three-momenta and quantum time coordinate are somewhat related.
In fact, for a minimum-uncertainty state
in the framework of $\kappa$-deformed kinematics
one has $\Delta  E \, \Delta  t \sim \hbar/2$ and
$\Delta  p \, \Delta  t \sim l_\kappa p/(2 c)$,
and this is consistent with a given dispersion relation $E(p)$
only if  $E(p) \sim (c \hbar/ l_\kappa) \ln (p/p^*)$
(with $p^*$ a constant to be otherwise determined)
which coincides with the asymptotic behavior of
the $\kappa$-deformed dispersion relation (see (\ref{2.19})).}
(\ref{2.20b}) one
finds that\begin{equation}\label{deltavofE}
\Delta	v \sim	\frac{\Delta  P}{\kappa}
\sim \frac{\hbar}{2 \kappa \, \Delta x}\,,
\end{equation}
where on the right-hand-side we used (2.15b).

Using (\ref{3starb}) and (\ref{deltavofE})
one can rewrite (\ref{3.1light}) as
\begin{equation}\label{3star}
\Delta L \geq
l_p \sqrt{ c T \over s} +
\Delta	x +
\frac{\hbar L}{2 \kappa c\, \Delta  x} +
\frac{\hbar T}{2 \kappa \, \Delta  x}
~,
\end{equation}
and therefore, also taking into account that $L \sim c \, T/2$
and $l_\kappa \equiv \hbar/(\kappa c)$,
one finds that the minimal value of $\Delta  L$ is obtained if
$(\Delta x)^2 \sim  L l_\kappa$ and this implies that the
minimal uncertainty in the measurement of the distance $L$ is
\begin{equation}\label{3.6}
\min[\Delta L] \sim \sqrt{ L l_p^2 \over s} + \sqrt{ L l_\kappa}
\end{equation}
Again we find the $\sqrt{L}$ behavior,
and again the full structure of the
covariantly $\kappa$-deformed phase space
advocated here
plays a rather central role in obtaining this result;
in fact, the relation (2.18) ensures that the fourth term on the
right-hand side of Eq.~(\ref{3.1light}) (which was not considered
in Ref.~\cite{kpoinpap})
is of the same order
as the second term on the
right-hand side of Eq.~(\ref{3.1light}), which is the one
considered in Ref.~\cite{kpoinpap}.

While the $\sqrt{L}$ behavior
is of course the most robust outcome of these analyses,
it is interesting to notice the interplay
between the scale $l_\kappa$,
which characterizes the $\kappa$ deformation,
and the scales $s$ and $l_p$,
which characterize heuristic quantum-gravity arguments.
Although the relative magnitude of these scales
could only be determined once a full Quantum Gravity
formalism became available,
it is quite natural
to guess that if $\kappa$ deformations were to have physical applications
it might be that $l_\kappa \sim l_p$. Moreover, from the role of $s$
in the measurement procedure it is clear \cite{gacmpla,kpoinpap}
that $s \ge l_p$, and since the measurability bound
should be a general property of the theory
it is conceivable that also $s \sim l_p$.
This, for example, appears to fit rather well the schemes,
such as the one discussed in Ref.~\cite{rovelli},
in which ``fundamental clocks'' are intrinsic to the formulation
of the quantum-gravity approach.
For $l_\kappa \sim l_p \sim s$ the heavy probe and the massless probe
considered in this and in the previous subsection lead to exactly
(up to an overall numerical factor of order 1)
the same bound in the context of the Wigner measurement,
and even the heuristic quantum-gravity measurement analysis
of Ref.~\cite{gacmpla} reproduces this bound exactly
(again up to an overall numerical factor of order 1).
Nevertheless, especially in light of the fact
that very little will be known
about $s$ until a fully consistent (and genuinely quantum) theory
of gravity is available,
it is interesting to observe that if $s \ne l_p$ ({\it i.e.} $s > l_p$)
the Wigner measurement using a heavy probe is actually
a ``better measurement'' (weaker bound) than its counterpart using
a massless probe.
Since most of the previous studies of quantum-gravity measurability
bounds have relied on massless probes, our results suggest
that a reanalysis of those studies might be necessary.

\section{Closing Remarks}

The covariant $\kappa$-deformation of relativistic symmetries
here considered, and the associated
covariant $\kappa$-deformation of the
Heisenberg algebra (2.14),
has several appealing properties as a candidate for the
high-energy modification of classical relativistic symmetries.
It provides
a rather moderate (at least in comparison
with some of its alternatives) deformation of classical
relativistic symmetries, which in particular reflects the
reasonable expectation that, if any of the space-time
coordinates is to be special, the special coordinate should be time.
(Interestingly this intuition appears to be also realized
in certain approaches to string theory, see {\it e.g.} Ref.~\cite{emn}.)
As manifest in the
relations (2.15a)-(2.15d),
the $\kappa$-modifications of the covariant
Heisenberg commutations relations are of quantum mechanical nature, i.e.
proportional to the Planck constant $\hbar$. This suggests that the
$\kappa$-deformation 
can be related with the quantum corrections to the
classical dynamics of the space-time geometry.
In extending the analysis of Ref.~\cite{kpoinpap}
from the space-time coordinate sector
to the full structure of the $\kappa$-deformed phase space,
our analysis has provided additional evidence that the bounds
on the measurability of distances associated with the
uncertainty relations characterizing the $\kappa$-deformed
covariant Heisenberg algebra (2.14) are consistent with
independent heuristic
quantum-gravity analyses of such measurability bounds.
Based on this consistency between heuristic quantum-gravity measurability
analysis and $\kappa$-Poincar\'e measurability analysis
one is prompted to consider the possibility
that \cite{kpoinpap,grf98ess}
at length scales larger than the Planck length
(but of course smaller than the length scales already
probed experimentally)
the $\kappa$-deformations of Poincar\'e symmetries might play a role
in the effective description of Quantum Gravity.
This is not completely surprising since some of the 
most popular Quantum-Gravity scenarios, such as the ones based on
a spacetime lattice and the ones involving a foamy Quantum-Gravity vacuum,
would plausibly lead \cite{grf98ess}
to deformations of Poincar\'e symmetries,
and, in particular, the analysis of Ref.~\cite{aemn1} appears
to suggest that propagation in a foamy Quantum Gravity vacuum
might be characterized by a $\kappa$-deformed dispersion relation.

The three-momentum-dependent (i.e. energy-dependent)
``speed of light'' (\ref{2.20b}) is a novel phenomenon that arises
in the framework here considered. 
As mentioned, it has the same
functional form (upon appropriate identification between
$\kappa$ and the string scale) as the energy-dependent speed of light
recently discussed \cite{aemn1}
in the non-critical  (``Liouville'') string  literature.
Both in the $\kappa$-Poincar\'{e} and in the string theory
contexts the deviation from ordinary physics,
while very significant at the conceptual level,
is rather marginal from the phenomenological viewpoint.
For example, for photons of energies of order 1 GeV
the Eq.~(\ref{2.20b}) entails a
minuscule $10^{-19} c$ correction with respect to the ordinary
scenario with constant speed of light.
At least when
$\kappa$ is identified with the Planck scale, the Eq.~(\ref{2.20b})
is completely consistent with presently available experimental
data \cite{aemn1,domo,bowjar}.
However (as already emphasized in Refs.~\cite{grbgac,glalorentz} 
and references therein), some of the modern techniques of 
investigation of astrophysical phenomena could soon bring remarkable
progress in the investigation of space-time symmetries. 
In our case the best laboratory
appears to be provided by gamma-ray bursts, and,
now that there is convincing evidence \cite{grbnew}
that these bursts have cosmological origin,
we can expect \cite{grbgac} that within a few years 
gamma-ray-burst data will test conclusively the $\kappa$ deformations
we considered.

\end{document}